 % EXTERNAL FILES

\input harvmac
\input amssym

\def\S{{\bf S}}

\def\Z{{\bf Z}}
\def\R{{\bf R}}

\def\K3{{\bf K3}}
\def\journal#1&#2(#3){\unskip, \sl #1\ \bf #2 \rm(19#3) }
\def\andjournal#1&#2(#3){\sl #1~\bf #2 \rm (19#3) }

\def\bar{\overline}
\def\hat{\widehat}

\def\frac#1#2{{#1\over#2}}

\def\inbar{\,\vrule height1.5ex width.4pt depth0pt}
\def\IC{\relax\hbox{$\inbar\kern-.3em{\rm C}$}}
\def\IR{\relax{\rm I\kern-.18em R}}
\def\IP{\relax{\rm I\kern-.18em P}}
\def\Z{{\bf Z}}

%
%%%%%%%%%%%%%%%%%%%%%%%%%%%%%%%%%%%%
%

%
\catcode`\@=11
\def\slash#1{\mathord{\mathpalette\c@ncel{#1}}}
\overfullrule=0pt

\def\underrel#1\over#2{\mathrel{\mathop{\kern\z@#1}\limits_{#2}}}

\catcode`\@=12

%%%%%%%%%%%%%%%%%%%%%%%%%%%%%%%%%%%%%%%%%%%%%%%%%%%%%%%%%%%%%%

%

\def\exp{{\rm exp}}

\def\unit{\relax{\rm 1\kern-.26em I}}
\def\nada{\relax{\rm 0\kern-.30em l}}

% \draftmode

%\def\Omega{\rho,\sigma,\nu  }

\def\CP{{\cal P}}
%% MACROS
\noblackbox
\def\IL{\relax{\rm I\kern-.18em L}}
\def\IH{\relax{\rm I\kern-.18em H}}
\def\IR{\relax{\rm I\kern-.18em R}}
\def\IC{\relax\hbox{$\inbar\kern-.3em{\rm C}$}}
\def\IZ{\relax\ifmmode\mathchoice
{\hbox{\cmss Z\kern-.4em Z}}{\hbox{\cmss Z\kern-.4em Z}} {\lower.9pt\hbox{\cmsss Z\kern-.4em Z}}
{\lower1.2pt\hbox{\cmsss Z\kern-.4em Z}}\else{\cmss Z\kern-.4em Z}\fi}

\def\CN {{\cal N}}

\def\CP {{\cal P }}
\def\CL {{\cal L}}

\def\CO {{\cal O}}

%% MORE MACROS

\def\CN {{\cal N}}

\def\CO {{\cal O}}

\def\CP {{\cal P }}

\font\manual=manfnt \def\dbend{\lower3.5pt\hbox{\manual\char127}}

\def\IZ{\relax\ifmmode\mathchoice
{\hbox{\cmss Z\kern-.4em Z}}{\hbox{\cmss Z\kern-.4em Z}} {\lower.9pt\hbox{\cmsss Z\kern-.4em Z}}
{\lower1.2pt\hbox{\cmsss Z\kern-.4em Z}}\else{\cmss Z\kern-.4em Z}\fi}

\def\bar{\overline}

\def\pa{\partial}

\def\rt2{\sqrt{2}}
\def\irt2{{1\over\sqrt{2}}}

\def\hat{\widehat}
%  \slashchar puts a slash through a character to represent contraction
%  with Dirac matrices. Use \not instead for negation of relations, and use
%  \hbar for hbar.
\def\slashchar#1{\setbox0=\hbox{$#1$}           % set a box for #1
   \dimen0=\wd0                                 % and get its size
   \setbox1=\hbox{/} \dimen1=\wd1               % get size of /
   \ifdim\dimen0>\dimen1                        % #1 is bigger
      \rlap{\hbox to \dimen0{\hfil/\hfil}}      % so center / in box
      #1                                        % and print #1
   \else                                        % / is bigger
      \rlap{\hbox to \dimen1{\hfil$#1$\hfil}}   % so center #1
      /                                         % and print /
   \fi}

\def\foursqr#1#2{{\vcenter{\vbox{
    \hrule height.#2pt
    \hbox{\vrule width.#2pt height#1pt \kern#1pt
    \vrule width.#2pt}
    \hrule height.#2pt
    \hrule height.#2pt
    \hbox{\vrule width.#2pt height#1pt \kern#1pt
    \vrule width.#2pt}
    \hrule height.#2pt
        \hrule height.#2pt
    \hbox{\vrule width.#2pt height#1pt \kern#1pt
    \vrule width.#2pt}
    \hrule height.#2pt
        \hrule height.#2pt
    \hbox{\vrule width.#2pt height#1pt \kern#1pt
    \vrule width.#2pt}
    \hrule height.#2pt}}}}
\def\psqr#1#2{{\vcenter{\vbox{\hrule height.#2pt
    \hbox{\vrule width.#2pt height#1pt \kern#1pt
    \vrule width.#2pt}
    \hrule height.#2pt \hrule height.#2pt
    \hbox{\vrule width.#2pt height#1pt \kern#1pt
    \vrule width.#2pt}
    \hrule height.#2pt}}}}
\def\sqr#1#2{{\vcenter{\vbox{\hrule height.#2pt
    \hbox{\vrule width.#2pt height#1pt \kern#1pt
    \vrule width.#2pt}
    \hrule height.#2pt}}}}

\def\figin{\epsfcheck\figin}\def\figins{\epsfcheck\figins}
\def\epsfcheck{\ifx\epsfbox\UnDeFiNeD
\message{(NO epsf.tex, FIGURES WILL BE IGNORED)}
\gdef\figin##1{\vskip2in}\gdef\figins##1{\hskip.5in}% blank space instead
\else\message{(FIGURES WILL BE INCLUDED)}%
\gdef\figin##1{##1}\gdef\figins##1{##1}\fi}
\def\DefWarn#1{}
\def\figinsert{\goodbreak\midinsert}
\def\ifig#1#2#3{\DefWarn#1\xdef#1{fig.~\the\figno}
\writedef{#1\leftbracket fig.\noexpand~\the\figno}%
\figinsert\figin{\centerline{#3}}\medskip\centerline{\vbox{\baselineskip12pt \advance\hsize by
-1truein\noindent\footnotefont{\bf Fig.~\the\figno:\ } \it#2}}
\bigskip\endinsert\global\advance\figno by1}

%%%%%%%%%%%%%%%%%%%%%%%%%%%%%%%%%%%%%%%%%%%%%%%%%%%%%%%%%%%%%%
% new defs:

% FONTS

% fraktur

\newfam\frakfam
\font\teneufm=eufm10
\font\seveneufm=eufm7
\font\fiveeufm=eufm5
\textfont\frakfam=\teneufm
\scriptfont\frakfam=\seveneufm
\scriptscriptfont\frakfam=\fiveeufm

% black board bold

\def\bb{
\font\tenmsb=msbm10
\font\sevenmsb=msbm7
\font\fivemsb=msbm5
\textfont1=\tenmsb
\scriptfont1=\sevenmsb
\scriptscriptfont1=\fivemsb
}

%\newfam\msbfam
%\font\tenmsb=msbm10
%\font\sevenmsb=msbm7
%\font\fivemsb=msbm5
%\textfont\msbfam=\tenmsb
%\scriptfont\msbfam=\sevenmsb
%\scriptscriptfont\msbfam=\fivemsb
%\def\bb{\fam\msbfam \tenmsb}

% double stroke math

\newfam\dsromfam
\font\tendsrom=dsrom10
\textfont\dsromfam=\tendsrom
\def\ds{\fam\dsromfam \tendsrom}

% bold math italics

\newfam\mbffam
\font\tenmbf=cmmib10
\font\sevenmbf=cmmib7
\font\fivembf=cmmib5
\textfont\mbffam=\tenmbf
\scriptfont\mbffam=\sevenmbf
\scriptscriptfont\mbffam=\fivembf

% bold math cal

\newfam\mbfcalfam
\font\tenmbfcal=cmbsy10
\font\sevenmbfcal=cmbsy7
\font\fivembfcal=cmbsy5
\textfont\mbfcalfam=\tenmbfcal
\scriptfont\mbfcalfam=\sevenmbfcal
\scriptscriptfont\mbfcalfam=\fivembfcal

% math script

\newfam\mscrfam
\font\tenmscr=rsfs10
\font\sevenmscr=rsfs7
\font\fivemscr=rsfs5
\textfont\mscrfam=\tenmscr
\scriptfont\mscrfam=\sevenmscr
\scriptscriptfont\mscrfam=\fivemscr

% MACROS

% bras, kets, ...

% tilde, hat, bar, ...

\def\hat{\widehat}

\def\bar{\overline}
\def\b{\bar}
\def\bsq#1{{{\b{#1}}^{\lower 2.5pt\hbox{$\scriptstyle 2$}}}}
\def\bexp#1#2{{{\b{#1}}^{\lower 2.5pt\hbox{$\scriptstyle #2$}}}}
\def\dotexp#1#2{{{#1}^{\lower 2.5pt\hbox{$\scriptstyle #2$}}}}

% basic math

\def\rt2{\sqrt{2}}

% bold greek characters

\font\tenbifull=cmmib10
\font\tenbimed=cmmib7
\font\tenbismall=cmmib5
\textfont9=\tenbifull \scriptfont9=\tenbimed
\scriptscriptfont9=\tenbismall

\mathchardef\bbGamma="7000
\mathchardef\bbDelta="7001
\mathchardef\bbPhi="7002
\mathchardef\bbAlpha="7003
\mathchardef\bbXi="7004
\mathchardef\bbPi="7005
\mathchardef\bbSigma="7006
\mathchardef\bbUpsilon="7007
\mathchardef\bbTheta="7008
\mathchardef\bbPsi="7009
\mathchardef\bbOmega="700A
\mathchardef\bbalpha="710B
\mathchardef\bbbeta="710C
\mathchardef\bbgamma="710D
\mathchardef\bbdelta="710E
\mathchardef\bbepsilon="710F
\mathchardef\bbzeta="7110
\mathchardef\bbeta="7111
\mathchardef\bbtheta="7112
\mathchardef\bbiota="7113
\mathchardef\bbkappa="7114
\mathchardef\bblambda="7115
\mathchardef\bbmu="7116
\mathchardef\bbnu="7117
\mathchardef\bbxi="7118
\mathchardef\bbpi="7119
\mathchardef\bbrho="711A
\mathchardef\bbsigma="711B
\mathchardef\bbtau="711C
\mathchardef\bbupsilon="711D
\mathchardef\bbphi="711E
\mathchardef\bbchi="711F
\mathchardef\bbpsi="7120
\mathchardef\bbomega="7121
\mathchardef\bbvarepsilon="7122
\mathchardef\bbvartheta="7123
\mathchardef\bbvarpi="7124
\mathchardef\bbvarrho="7125
\mathchardef\bbvarsigma="7126
\mathchardef\bbvarphi="7127

% dotted spinor indices

% bared indices

% bared spinors

% capital cal letters

\def\CL{{\cal L}}

\def\CN{{\cal N}}
\def\CO{{\cal O}}
\def\CP{{\cal P}}

% double stroke symbols: unit matrix, reals, complex, quaternions, integers, natural numbers

\def\1{{\ds 1}}
\def\R{\hbox{$\bb R$}}

\def\Z{\hbox{$\bb Z$}}
\def\S{\hbox{$\bb S$}}

\def\CP{\hbox{$\bb CP$}}

% miscellaneous objects

%\SohniusTP
\lref\SohniusTP{
  M.~F.~Sohnius and P.~C.~West,
  ``An Alternative Minimal Off-Shell Version Of N=1 Supergravity,''
  Phys.\ Lett.\  B {\bf 105}, 353 (1981).
  %%CITATION = PHLTA,B105,353;%%
}

%\GatesNR
\lref\GatesNR{
  S.~J.~Gates, M.~T.~Grisaru, M.~Rocek and W.~Siegel,
  ``Superspace, or one thousand and one lessons in supersymmetry,''
  Front.\ Phys.\  {\bf 58}, 1 (1983)
  [arXiv:hep-th/0108200].
  %%CITATION = FRPHA,58,1;%%
}

%\KomargodskiPC
\lref\KomargodskiPC{
  Z.~Komargodski and N.~Seiberg,
  ``Comments on the Fayet-Iliopoulos Term in Field Theory and Supergravity,''
  JHEP {\bf 0906}, 007 (2009)
  [arXiv:0904.1159 [hep-th]].
  %%CITATION = JHEPA,0906,007;%%
}

%\ArkaniHamedRS
\lref\ArkaniHamedRS{
  N.~Arkani-Hamed, S.~Dimopoulos and G.~R.~Dvali,
  ``The hierarchy problem and new dimensions at a millimeter,''
  Phys.\ Lett.\  B {\bf 429}, 263 (1998)
  [arXiv:hep-ph/9803315].
  %%CITATION = PHLTA,B429,263;%%
}

%\LambertDX
\lref\LambertDX{
  N.~D.~Lambert and G.~W.~Moore,
  ``Distinguishing off-shell supergravities with on-shell physics,''
  Phys.\ Rev.\  D {\bf 72}, 085018 (2005)
  [arXiv:hep-th/0507018].
  %%CITATION = PHRVA,D72,085018;%%
}

%\AkulovCK
\lref\AkulovCK{
  V.~P.~Akulov, D.~V.~Volkov and V.~A.~Soroka,
  ``On The General Covariant Theory Of Calibrating Poles In Superspace,''
  Theor.\ Math.\ Phys.\  {\bf 31}, 285 (1977)
  [Teor.\ Mat.\ Fiz.\  {\bf 31}, 12 (1977)].
  %%CITATION = TMFZA,31,12;%%
}

%\KalloshVE
\lref\KalloshVE{
  R.~Kallosh, L.~Kofman, A.~D.~Linde and A.~Van Proeyen,
  ``Superconformal symmetry, supergravity and cosmology,''
  Class.\ Quant.\ Grav.\  {\bf 17}, 4269 (2000)
  [Erratum-ibid.\  {\bf 21}, 5017 (2004)]
  [arXiv:hep-th/0006179].
  %%CITATION = CQGRD,17,4269;%%
}

%\DvaliZH
\lref\DvaliZH{
  G.~Dvali, R.~Kallosh and A.~Van Proeyen,
  ``D-term strings,''
  JHEP {\bf 0401}, 035 (2004)
  [arXiv:hep-th/0312005].
  %%CITATION = JHEPA,0401,035;%%
}

%\ColemanUZ
\lref\ColemanUZ{
  S.~R.~Coleman,
  ``More About The Massive Schwinger Model,''
  Annals Phys.\  {\bf 101}, 239 (1976).
  %%CITATION = APNYA,101,239;%%
}

%\FischlerZK
\lref\FischlerZK{
  W.~Fischler, H.~P.~Nilles, J.~Polchinski, S.~Raby and L.~Susskind,
  ``Vanishing Renormalization Of The D Term In Supersymmetric U(1) Theories,''
  Phys.\ Rev.\ Lett.\  {\bf 47}, 757 (1981).
  %%CITATION = PRLTA,47,757;%%
}

%\WittenNF
\lref\WittenNF{
  E.~Witten,
  ``Dynamical Breaking Of Supersymmetry,''
  Nucl.\ Phys.\  B {\bf 188}, 513 (1981).
  %%CITATION = NUPHA,B188,513;%%
}

\lref\Wittenun{E.~Witten, unpublished}

%\Distlerun
\lref\Distlerun{
  J.~Distler and B.~Wecht,
  Unpublished, mentioned in {\sl http://golem.ph.utexas.edu/~distler
  /blog/archives/002180.html  }
  }
%\GreeneYA
\lref\GreeneYA{
  B.~R.~Greene, A.~D.~Shapere, C.~Vafa and S.~T.~Yau,
  ``Stringy Cosmic Strings And Noncompact Calabi-Yau Manifolds,''
  Nucl.\ Phys.\  B {\bf 337}, 1 (1990).
  %%CITATION = NUPHA,B337,1;%%
}
%\AshokGK
\lref\AshokGK{
  S.~Ashok and M.~R.~Douglas,
  ``Counting flux vacua,''
  JHEP {\bf 0401}, 060 (2004)
  [arXiv:hep-th/0307049].
  %%CITATION = JHEPA,0401,060;%%
}

%\ElvangJK
\lref\ElvangJK{
  H.~Elvang, D.~Z.~Freedman and B.~Kors,
  ``Anomaly cancellation in supergravity with Fayet-Iliopoulos couplings,''
  JHEP {\bf 0611}, 068 (2006)
  [arXiv:hep-th/0606012].
  %%CITATION = JHEPA,0611,068;%%
}

%\ShifmanZI
\lref\ShifmanZI{
  M.~A.~Shifman and A.~I.~Vainshtein,
  ``Solution of the Anomaly Puzzle in SUSY Gauge Theories and the Wilson
  Operator Expansion,''
  Nucl.\ Phys.\  B {\bf 277}, 456 (1986)
  [Sov.\ Phys.\ JETP {\bf 64}, 428 (1986\ ZETFA,91,723-744.1986)].
  %%CITATION = ZETFA,91,723;%%
}

%\BarbieriAC
\lref\BarbieriAC{
  R.~Barbieri, S.~Ferrara, D.~V.~Nanopoulos and K.~S.~Stelle,
  ``Supergravity, R Invariance And Spontaneous Supersymmetry Breaking,''
  Phys.\ Lett.\  B {\bf 113}, 219 (1982).
  %%CITATION = PHLTA,B113,219;%%
}

%\SiegelSV
\lref\SiegelSV{
  W.~Siegel,
  ``16/16 Supergravity,''
  Class.\ Quant.\ Grav.\  {\bf 3}, L47 (1986).
  %%CITATION = CQGRD,3,L47;%%
}

%\KuzenkoAM
\lref\KuzenkoAM{
  S.~M.~Kuzenko,
  %``Variant supercurrent multiplets,''
  arXiv:1002.4932 [hep-th].
  %%CITATION = ARXIV:1002.4932;%%
}

%\GirardiVQ
\lref\GirardiVQ{
  G.~Girardi, R.~Grimm, M.~Muller and J.~Wess,
  ``Antisymmetric Tensor Gauge Potential In Curved Superspace And A (16+16)
  Supergravity Multiplet,''
  Phys.\ Lett.\  B {\bf 147}, 81 (1984).
  %%CITATION = PHLTA,B147,81;%%
}

%\LangXK
\lref\LangXK{
  W.~Lang, J.~Louis and B.~A.~Ovrut,
  ``(16+16) Supergravity Coupled To Matter: The Low-Energy Limit Of The
  Superstring,''
  Phys.\ Lett.\  B {\bf 158}, 40 (1985).
  %%CITATION = PHLTA,B158,40;%%
}

%\CaldararuTC
\lref\CaldararuTC{
  A.~Caldararu, J.~Distler, S.~Hellerman, T.~Pantev and E.~Sharpe,
  ``Non-birational twisted derived equivalences in abelian GLSMs,''
  arXiv:0709.3855 [hep-th].
  %%CITATION = ARXIV:0709.3855;%%
}

%\SeibergVC
\lref\SeibergVC{
  N.~Seiberg,
  ``Naturalness Versus Supersymmetric Non-renormalization Theorems,''
  Phys.\ Lett.\  B {\bf 318}, 469 (1993)
  [arXiv:hep-ph/9309335].
  %%CITATION = PHLTA,B318,469;%%
}

%\DineXK
\lref\DineXK{
  M.~Dine, N.~Seiberg and E.~Witten,
  ``Fayet-Iliopoulos Terms in String Theory,''
  Nucl.\ Phys.\  B {\bf 289}, 589 (1987).
  %%CITATION = NUPHA,B289,589;%%
}

%\FreedmanUK
\lref\FreedmanUK{
  D.~Z.~Freedman,
  ``Supergravity With Axial Gauge Invariance,''
  Phys.\ Rev.\  D {\bf 15}, 1173 (1977).
  %%CITATION = PHRVA,D15,1173;%%
}

%\DasPU
\lref\DasPU{
  A.~Das, M.~Fischler and M.~Rocek,
  ``Superhiggs Effect In A New Class Of Scalar Models And A Model Of Super
  QED,''
  Phys.\ Rev.\  D {\bf 16}, 3427 (1977).
  %%CITATION = PHRVA,D16,3427;%%
}

     %\deWitWW
      \lref\deWitWW{
        B.~de Wit and P.~van Nieuwenhuizen,
       ``The Auxiliary Field Structure In Chirally Extended Supergravity,''
        Nucl.\ Phys.\  B {\bf 139}, 216 (1978).
        %%CITATION = NUPHA,B139,216;%%
      }

%\BinetruyHH
\lref\BinetruyHH{
  P.~Binetruy, G.~Dvali, R.~Kallosh and A.~Van Proeyen,
  ``Fayet-Iliopoulos terms in supergravity and cosmology,''
  Class.\ Quant.\ Grav.\  {\bf 21}, 3137 (2004)
  [arXiv:hep-th/0402046].
  %%CITATION = CQGRD,21,3137;%%
}

%\FerraraDH
\lref\FerraraDH{
  S.~Ferrara, L.~Girardello, T.~Kugo and A.~Van Proeyen,
  ``Relation Between Different Auxiliary Field Formulations Of N=1 Supergravity
  Coupled To Matter,''
  Nucl.\ Phys.\  B {\bf 223}, 191 (1983).
  %%CITATION = NUPHA,B223,191;%%
}

%\WeinbergUV
\lref\WeinbergUV{
  S.~Weinberg,
  ``Non-renormalization theorems in non-renormalizable theories,''
  Phys.\ Rev.\ Lett.\  {\bf 80}, 3702 (1998)
  [arXiv:hep-th/9803099].
  %%CITATION = PRLTA,80,3702;%%
}

%\FerraraPZ
\lref\FerraraPZ{
  S.~Ferrara and B.~Zumino,
  ``Transformation Properties Of The Supercurrent,''
  Nucl.\ Phys.\  B {\bf 87}, 207 (1975).
  %%CITATION = NUPHA,B87,207;%%
}

%\IntriligatorCP
\lref\IntriligatorCP{
  K.~A.~Intriligator and N.~Seiberg,
  ``Lectures on Supersymmetry Breaking,''
  Class.\ Quant.\ Grav.\  {\bf 24}, S741 (2007)
  [arXiv:hep-ph/0702069].
  %%CITATION = CQGRD,24,S741;%%
}

%\DineTA
\lref\DineTA{
  M.~Dine,
  ``Fields, Strings and Duality: TASI 96,''
 eds. C.~Efthimiou and B. Greene (World Scientific, Singapore, 1997).
   }

%\FerraraPZ
\lref\FerraraPZ{
  S.~Ferrara and B.~Zumino,
  ``Transformation Properties Of The Supercurrent,''
  Nucl.\ Phys.\  B {\bf 87}, 207 (1975).
  %%CITATION = NUPHA,B87,207;%%
}

%\WittenBZ
\lref\WittenBZ{
  E.~Witten,
  ``New Issues In Manifolds Of SU(3) Holonomy,''
  Nucl.\ Phys.\  B {\bf 268}, 79 (1986).
  %%CITATION = NUPHA,B268,79;%%
}

%\HorowitzNG
\lref\HorowitzNG{
  G.~T.~Horowitz,
  ``Exactly Soluble Diffeomorphism Invariant Theories,''
  Commun.\ Math.\ Phys.\  {\bf 125}, 417 (1989).
  %%CITATION = CMPHA,125,417;%%
}

%\SeibergVC
\lref\SeibergVC{
  N.~Seiberg,
 ``Naturalness Versus Supersymmetric Non-renormalization Theorems,''
  Phys.\ Lett.\  B {\bf 318}, 469 (1993)
  [arXiv:hep-ph/9309335].
  %%CITATION = PHLTA,B318,469;%%
}

%\O'RaifeartaighPR
\lref\ORaifeartaighPR{
  L.~O'Raifeartaigh,
  ``Spontaneous Symmetry Breaking For Chiral Scalar Superfields,''
  Nucl.\ Phys.\  B {\bf 96}, 331 (1975).
  %%CITATION = NUPHA,B96,331;%%
}

%\FayetJB
\lref\FayetJB{
  P.~Fayet and J.~Iliopoulos,
  ``Spontaneously Broken Supergauge Symmetries and Goldstone Spinors,''
  Phys.\ Lett.\  B {\bf 51}, 461 (1974).
  %%CITATION = PHLTA,B51,461;%%
}

%\GreenSG
\lref\GreenSG{
  M.~B.~Green and J.~H.~Schwarz,
  ``Anomaly Cancellation In Supersymmetric D=10 Gauge Theory And Superstring
  Theory,''
  Phys.\ Lett.\  B {\bf 149}, 117 (1984).
  %%CITATION = PHLTA,B149,117;%%
}

%\PantevRH
\lref\PantevRH{
  T.~Pantev and E.~Sharpe,
  ``Notes on gauging noneffective group actions,''
  arXiv:hep-th/0502027.
  %%CITATION = HEP-TH/0502027;%%
}

%\PantevZS
\lref\PantevZS{
  T.~Pantev and E.~Sharpe,
  ``GLSM's for gerbes (and other toric stacks),''
  Adv.\ Theor.\ Math.\ Phys.\  {\bf 10}, 77 (2006)
  [arXiv:hep-th/0502053].
  %%CITATION = 00203,10,77;%%
}

%\ChamseddineGB
\lref\ChamseddineGB{
  A.~H.~Chamseddine and H.~K.~Dreiner,
  ``Anomaly Free Gauged R Symmetry In Local Supersymmetry,''
  Nucl.\ Phys.\  B {\bf 458}, 65 (1996)
  [arXiv:hep-ph/9504337].
  %%CITATION = NUPHA,B458,65;%%
}

%\CastanoCI
\lref\CastanoCI{
  D.~J.~Castano, D.~Z.~Freedman and C.~Manuel,
  ``Consequences of supergravity with gauged U(1)-R symmetry,''
  Nucl.\ Phys.\  B {\bf 461}, 50 (1996)
  [arXiv:hep-ph/9507397].
  %%CITATION = NUPHA,B461,50;%%
}

%\WittenHU
\lref\WittenHU{
  E.~Witten and J.~Bagger,
  ``Quantization Of Newton's Constant In Certain Supergravity Theories,''
  Phys.\ Lett.\  B {\bf 115}, 202 (1982).
  %%CITATION = PHLTA,B115,202;%%
}

%\BaggerFN
\lref\BaggerFN{
  J.~Bagger and E.~Witten,
  ``The Gauge Invariant Supersymmetric Nonlinear Sigma Model,''
  Phys.\ Lett.\  B {\bf 118}, 103 (1982).
  %%CITATION = PHLTA,B118,103;%%
}

%\BanksMB
\lref\BanksMB{
  T.~Banks, M.~Dine and N.~Seiberg,
  ``Irrational axions as a solution of the strong CP problem in an eternal
  universe,''
  Phys.\ Lett.\  B {\bf 273}, 105 (1991)
  [arXiv:hep-th/9109040].
  %%CITATION = PHLTA,B273,105;%%
}

%\KomargodskiRB
\lref\KomargodskiRB{
  Z.~Komargodski and N.~Seiberg,
  ``Comments on Supercurrent Multiplets, Supersymmetric Field Theories and
  Supergravity,''
  arXiv:1002.2228 [hep-th].
  %%CITATION = ARXIV:1002.2228;%%
}

%\KuzenkoAM
\lref\KuzenkoAM{
  S.~M.~Kuzenko,
  ``Variant supercurrent multiplets,''
  arXiv:1002.4932 [hep-th].
  %%CITATION = ARXIV:1002.4932;%%
}

%\DienesTD
\lref\DienesTD{
  K.~R.~Dienes and B.~Thomas,
  ``On the Inconsistency of Fayet-Iliopoulos Terms in Supergravity Theories,''
  arXiv:0911.0677 [hep-th].
  %%CITATION = ARXIV:0911.0677;%%
}

%\KuzenkoYM
\lref\KuzenkoYM{
  S.~M.~Kuzenko,
  ``The Fayet-Iliopoulos term and nonlinear self-duality,''
  arXiv:0911.5190 [hep-th].
  %%CITATION = ARXIV:0911.5190;%%
}

%\WittenBZ
\lref\WittenBZ{
  E.~Witten,
  ``New Issues In Manifolds Of SU(3) Holonomy,''
  Nucl.\ Phys.\  B {\bf 268}, 79 (1986).
  %%CITATION = NUPHA,B268,79;%%
}

%\KugoFS
\lref\KugoFS{
  T.~Kugo and T.~T.~Yanagida,
  ``Coupling Supersymmetric Nonlinear Sigma Models to Supergravity,''
  arXiv:1003.5985 [hep-th].
  %%CITATION = ARXIV:1003.5985;%%
}

%%%%%%%%%%%%%%%%%%%%%%%%%%%%%%%%%%%%%%%%%%%%%%%%%%%
%\draftmode

\Title{
} {\vbox{\centerline{Modifying the Sum Over Topological Sectors}
\centerline{}
 \centerline{and}
 \centerline{}
 \centerline{Constraints on Supergravity}
}}
\medskip

\centerline{\it Nathan Seiberg }

\bigskip
\centerline{School of Natural Sciences}
\centerline{Institute for Advanced Study}
\centerline{Einstein Drive, Princeton, NJ 08540}

\smallskip

\vglue .3cm

\bigskip
\noindent
The standard lore about the sum over topological sectors in quantum field theory is that locality and cluster decomposition uniquely determine the sum over such sectors, thus leading to the usual $\theta$-vacua.  We show that without changing the local degrees of freedom, a theory can be modified such that the sum over instantons should be restricted; e.g.\ one should include only instanton numbers which are divisible by some integer $p$.  This conclusion about the configuration space of quantum field theory allows us to carefully reconsider the quantization of parameters in supergravity.  In particular, we show that FI-terms and nontrivial K\"ahler forms are quantized.  This analysis also leads to a new derivation of recent results about linearized supergravity.

\Date{April 2010}

\newsec{Introduction}

This note addresses two issues.  The first topic is purely field theoretic and the second topic involves supergravity.

Our field theory analysis was motivated by supergravity considerations.  We were interested in topological constraints on supergravity theories which lead to quantization of the parameters in the Lagrangian.  The proper understanding of the configuration space turns out to affect the correct quantization condition of these parameters.

For concreteness we will limit ourselves to $\CN=1$ supergravity in four dimensions.  We will find it convenient to distinguish two classes of theories:
\item{1.} Supersymmetric field theories coupled to supergravity.  Here we assume that the field theory has no parameter of order the Planck scale $M_{Planck}$ -- the K\"ahler potential and the superpotential are independent of $M_{Planck}$. All the dependence on the Planck scale arises either from the coupling to supergravity or from nonrenormalizable operators which have no effect on the low energy dynamics.  Here we exclude theories with moduli whose target space is of order the Planck scale.
\item{2.} Intrinsic supergravity theories.  Here some couplings of the non-gravitational fields are fixed to be of order $M_{Planck}$; i.e.\ they cannot be continuously varied.  In particular, these couplings cannot be parametrically smaller than $M_{Planck}$.  Therefore, such theories do not have a (rigid) field theory limit.

\medskip

In the first class of theories the field theory dynamics decouples from gravity and we can study supergravity theories by first analyzing the rigid limit.  The coupling to supergravity is determined in the linearized approximation by the energy momentum tensor and supercurrent of the rigid theory.

The investigation in \refs{\KomargodskiPC,\KomargodskiRB} was limited to theories in the first class.  A careful analysis of the supersymmetry current has shown that Abelian gauge theories (which include charged fields) with an FI-term $\xi$ and theories whose target space has non-exact K\"ahler form $\omega$ can be coupled to standard supergravity only if the theory has an exact continuous global symmetry.\foot{Such theories with FI-terms were originally studied in \refs{\FreedmanUK\BarbieriAC\KalloshVE-\DvaliZH} and more recently, following  \refs{\KomargodskiPC,\KomargodskiRB}, in \refs{\DienesTD\KuzenkoYM-\KuzenkoAM}.  To the best of our knowledge the analogous situation with nontrivial topology was not studied before \KomargodskiRB.  (The authors of \WittenHU\ studied theories in the second class.)}  However, as emphasized in \KomargodskiPC, the absence of global continuous symmetries in a gravitational theory makes such theories less interesting.

Reference \KomargodskiRB\ also considered field theories with nonzero $\xi$ or non-exact $\omega$.  Following standard gauging procedure of the supersymmetry current of these theories has led to a larger supergravity multiplet similar to the one in \refs{\GirardiVQ,\LangXK}.  As emphasized in~\SiegelSV\ and elaborated in~\KomargodskiRB, such theories can be reinterpreted as ordinary supergravity theories coupled to a modified matter system which has an additional chiral superfield.  The latter fixes the problems of the original field theory by making $\xi$ ``field dependent'' \DineXK\ and by ruining the topology underlying the non-exact $\omega$.

In this note we examine theories of the second class above in which $\xi \sim M_{Planck}^2$ and the target space has a nontrivial topology with $\int \omega \sim M_{Planck}^2$.  Such theories are inherently gravitational.

In section 2 we discuss a purely field theoretic problem.  In many field theories the configuration space splits into disconnected sectors labeled by the topological charge $n \in \Z$.  It is commonly stated that in order to satisfy locality and cluster decomposition one must sum over all these sectors with a weight factor $e^{in\theta}$.  In a Hamiltonian formalism this corresponds to considering $\theta$-vacua rather than $n$-vacua.  Section 2 argues that such theories can be deformed, without adding local degrees of freedom, such that the instanton sum must be modified.  For example, in some cases we should sum only over values of $n$ which are divisible by some integer $p$.

Section 3 is devoted to the constraints on supergravity theories in the second class which are intrinsically gravitational.  Here we argue that the FI-term is quantized\foot{We use notation ${1\over G_N} =M_{Planck}^2 = {8\pi \over \kappa^2 }= 8\pi M_P^2 $; i.e.\ $M_P$ is the reduced Planck mass.}
 \eqn\xicont{\xi = 2 N M_{P}^2    \qquad , \qquad N \in \Z ~. }
We also study theories with topologically nontrivial target spaces.  As in \WittenHU, the elements of the second cohomology group of the target space $H^2$ are constrained.  However, because of the subtleties discussed in section 2, these constraints are weaker than in \WittenHU.  Instead of studying the most general K\"ahler manifold, we focus on $\CP^1$ with the metric
\eqn\CPOmet{ds^2 = f_\pi^2 {d \Phi d \bar \Phi \over (1 + |\Phi|^2)^2}}
and we show that $f_\pi$ is constrained to satisfy
 \eqn\fpicon{f_\pi^2 = {2N \over  p} M_{P}^2  \qquad , \qquad p, N \in \Z ~.}
The integer parameter $p$ is the one we find in section 2.

We should emphasize that our discussion is incomplete for several reasons:
\item{1.}  We ignore the possible back-reaction of the spacetime metric. As pointed out in \refs{\GreeneYA,\AshokGK}, when topological objects of codimension two are present and a deficit angle in spacetime is generated, there can be additional constraints on the allowed parameters.
\item{2.} We focus on the classical theory.  Perturbative quantum considerations can lead to further restrictions like the requirement of anomaly cancelation.
\item{3.} Nonperturbative quantum effects are also important.  For example, the incompatibility of global continuous symmetries with gravity leads to additional conditions.
\item{4.} Finally, it is quite likely that there are other more subtle consistency conditions which we are not yet aware of.

In section 4 we study theories of the first class -- rigid supersymmetric field theories coupled to supergravity.  Here we use the results of section 3 to derive the known results about these theories.  In particular, we show that Abelian gauge theories with an arbitrary FI-term $\xi$ can be coupled to
supergravity only if the gauge group is noncompact; i.e.\ it is $\R$ rather than $U(1)$.  Furthermore, if the rigid theory includes charged fields, it should have a continuous global R-symmetry.  Section 4 also considers theories with a target space whose K\"ahler form $\omega$ is not exact and its periods are arbitrary.  Such theories can be coupled to minimal supergravity only if the theory has a continuous global R-symmetry and the total wrapping number of spacetime over the target space is constrained.  In these two situations of nonzero $\xi$ and arbitrary $\omega$ the resulting supergravity theory has an exact continuous global symmetry (and therefore such a theory is not expected to arise from a fully consistent theory of quantum gravity).

\newsec{The sum over topological sectors}

This section addresses the sum over topological sectors in quantum field theory.  Instead of presenting a general abstract theory, we will discuss simple examples.

\subsec{A trivial warmup}

As a warmup we review the situation in two dimensional Abelian gauge theories, emphasizing points which will be important below.

We start with the pure gauge $U(1)$ theory on a Euclidean compact spacetime.  The configuration space splits to ``instantons" labeled by the first Chern class
\eqn\instgau{{1\over 2\pi} \int F \in \Z ~.}
We are instructed to sum over these sectors with weight
\eqn\uoneth{e^{i {\theta \over 2\pi}\int F}}
and because of the quantization in \instgau, the result is periodic in $\theta$; i.e.\ $\theta \sim \theta + 2\pi$.  We can easily add charged particles to this system.  Their charges must be quantized.

The Hamiltonian interpretation of this system is obtained when spacetime is $\S^1\times \R$ and we view $\S^1$ as space and $\R$ as time.  As is well known, the parameter $\theta$ is interpreted as a background electric field \ColemanUZ. The Hamiltonian interpretation of these $\theta$-vacua involves two different elements which should not be confused:
\item{1.} The different values of $\theta$ in \uoneth\ label distinct superselection sectors.  Wilson line operators $\exp\left(i\oint A\right)$ where the integral is around our $\S^1$ space change the background electric field by one unit and shift $\theta$ by $2\pi$.  Hence the superselection sectors are labeled by  $-\pi < \theta \le \pi $.
\item{2.} A given superselection sector labeled by $\theta$ can include several stable states with different value of the background electric field.  Unlike the previous point which depends only on the configuration space, this is a more detailed issue, which depends also on the dynamical charges in the system and on the Hamiltonian.  If charge $p$ particles are present, the background electric field can be screened \ColemanUZ\ to be between $- {p\over 2}$ and $ p \over 2$.  This happens by creating a particle-antiparticle pair, moving one of them around the $\S^1$ space and then annihilating them.  Therefore, when $p \not=1$ each superselection sector labeled by $-\pi < \theta \le \pi$ includes $p$ stable states with different background electric field.  If there are several different charged particles with charges $p_i$, the stable values of the background electric field in each superselection sector are determined by their smallest common factor.

\medskip
If the gauge group is noncompact, the previous situation is modified.  On a compact spacetime the condition \instgau\ becomes $\int F=0$ and hence there is no $\theta $ parameter.  The Hamiltonian formalism interpretation of this fact is that the system does not have superselection sectors -- Wilson line operators $\exp\left(i r \oint A\right)$ with arbitrary real $r$ can set the background electric field to any value.  As above, the stability of states with background electric field is a more detailed question which depends on the dynamical charges.  If there are no dynamical charges, every state is stable.  If there are at least two charges whose ratio is irrational, every background electric field can be screened.

Finally, as emphasized, e.g\ in \BanksMB, if spacetime is taken to be $\R^2$, then the only notion of $\theta$-vacua is a stable state with background electric field.  This depends both on the gauge group and on the set of dynamical charges.

\subsec{The $\CP^1$ model}

Our second example is the two-dimensional $\CP^1$ sigma model.  The Lagrangian of the system is
 \eqn\sigmalag{\CL= f_\pi^2 {\partial_\mu\Phi\partial^\mu \bar \Phi \over (1+|\Phi|^2)^2}~.}
The target space is compactified to a $\CP^1$ by adding the point at infinity (which is at finite distance in the metric~\sigmalag). The patch around $\Phi=\infty$ is related to the other patch by the transformation
\eqn\Phitran{\Phi \to 1/\Phi~.}

We take our spacetime to be compact, e.g.\ $\S^2$.  Then the configuration space is divided into classes labeled by the wrapping number -- instanton number
\eqn\nunorm{I = \int \nu  \in \Z ~, }
where $\nu$ is proportional to the pull back of the K\"ahler form on $\CP^1$ to spacetime and is normalized such that \nunorm\ is satisfied.  Correspondingly, we can add to the Lagrangian \sigmalag\ a $\theta$ term
\eqn\thetaterb{i \theta \nu ~. }
Hence, $\theta$ has period $2\pi$.

It is often stated that we cannot restrict $I$ to any fixed value. This would amount to studying the ``$n$-vacua'' rather than the ``$\theta$-vacua'' and would be in conflict with cluster decomposition and locality.  Here we would like to reexamine this statement.

We add to the Lagrangian based on~\sigmalag\ and \thetaterb\
 \eqn\additer{\delta\CL= i \lambda \left( \nu-  {p\over 2\pi} F\right) + i {\hat \theta \over 2\pi} F}
which depends on the integer parameter $p$.  Here $\lambda(x)$ is a Lagrange multiplier and the two form $F$ is the field strength of a $U(1)$ gauge theory normalized as in \instgau.  Shifting $\lambda$ by a constant we can set $\hat \theta=0$ and obtain a new $\theta$ term $\theta\rightarrow\theta+\hat\theta/p$.

Has the addition of the terms \additer\ changed the theory?  The equation of motion of the gauge field of $F$ sets $\lambda$ to a constant. The Lagrange multiplier $\lambda$ sets
 \eqn\consla{\nu = {p \over 2\pi} F}
thus removing all the local degrees of freedom we added to the original $\CP^1$ model in \additer.

However, even though the original theory and the theory with \additer\ have the same set of local degrees of freedom, the two theories are actually different.  First, when our spacetime is compact, the constraint~\consla\ leads to
 \eqn\Iconst{I=\int \nu \in p\ \Z ~.}
For $p=1$ this condition is uninteresting.  For integer $p\not=1$ \Iconst\ states that the total instanton number must be a multiplet of $p$.\foot{If $p={m\over n} $ is rational (with $m$ and $n$ coprime),  $I=\int \nu \in m\ \Z$ and if $p$ is irrational, $I =\int\nu =0$.}  Therefore, $\theta$ in~\thetaterb\ has period $2\pi/p$.  Alternatively, we could use the freedom in shifting $\lambda$ to set $\theta=0$ and label the vacua by $\hat \theta$ in~\additer\ with period $2\pi$.  Second, when our spacetime is $\S^2$ we can solve \consla\ by setting the gauge field equal to the K\"ahler connection up to a gauge transformation.  But when our spacetime has nontrivial one cycles, e.g.\ when it is $\S^1 \times \R$ or a compact Riemann surface, the constraint \consla\ determines $F$ but leaves freedom in a nontrivial flat gauge field which should be integrated over.  Correspondingly, this theory has additional operators which the underlying sigma model does not have
\eqn\Wilop{W_m=e^{im \oint A} \qquad , \qquad m \in \Z }
where $A$ is the gauge field and the integral is over any nontrivial cycle.  Note that if the integral is over a topologically trivial cycle, or when $m$ is a multiple of $p$, the operator $W_m$ can be expressed in terms of the sigma model variable and was present before the theory was modified.

Considering the simple case where spacetime is $\S^2$, the partition function of the theory is related to the usual partition function of the $p=1$ theory as follows
\eqn\parfun{Z_{p}\left(\theta\right)=\sum_{n=0}^{p-1}Z_{p=1}\left(\theta+{2\pi n\over p}\right)~.}

In conclusion, the modified theory is locally the same as the original sigma model, but globally it is different.  In particular, the instanton sum is performed differently in the two theories.

We see that the standard lore about instantons in the $\CP^1$ theory corresponds to $p=1$.  However, for generic integer $p$ we find a theory with exactly the same local structure, but with the constraint~\Iconst\ on the total instanton number.  Since we added to the original Lagrangian \sigmalag\thetaterb\ the local term~\additer, it is clear that the resulting theory is local!

\medskip

We would like to make several comments:
\item{1.}  In the $p \to \infty$ limit the total instanton number $I=\int \nu $ must vanish.  One way to see it is by first rescaling the gauge field thus turning the $U(1)$ gauge theory into an $\R$ gauge theory in which $\int F =0$.  The same result is obtained for irrational $p$ in \additer.
\item{2.}  We are used to studying standard field theories based on local degrees of freedom in which the global structure does not matter much. We are also familiar with topological field theories which have no local degrees of freedom and whose entire dynamics depends on the global structure.  The theories we study here can be viewed as standard local theories coupled to topological theories.  In our example above the topological theory is a BF-theory \HorowitzNG, where the Lagrange multiplier $\lambda$ plays the role of the $B$ field and also couples to the sigma model variables.
\item{3.}  We can get further insight into the role of the gauge field in \additer\ by adding to the system a massive field $\varphi$ which couples to the gauge field with charge $q\in \Z $.  Working on $\S^2$ we can easily integrate out $\lambda$ and the gauge field to find that the massive field $\varphi$ couples to the massless mode $\Phi$.  For $p=1$ it couples to the K\"ahler connection of the $\CP^1$ and it ends up being a section of a line bundle on the $\CP^1$.  However, when $p\not=1$ and $q$ is not a multiple of $p$ such an interpretation is not possible.  Yet, when the constraint \Iconst\ is satisfied the massive field $\varphi$ is single valued and well defined on our $\S^2$ spacetime.
\item{4.}  We can add to the system several massive fields $\varphi$ with various charges and $U(1)$ gauge invariant interactions.  When the Lagrange multiplier $\lambda$ is integrated out, it eliminates the dynamical gauge field and the effective theory has a global $U(1)$ symmetry.  This global symmetry will play an important role below.
\item{5.}  As in our warmup discussion in section 2.1, the Hamiltonian interpretation of this setup is as follows.  We study the system on $\S^1 \times \R$ and view $\S^1$ as space and $\R$ as time.  The new operators \Wilop\ which wind around our space change $\hat \theta$ by $2\pi m$, or equivalently, they change $\theta$ of the underlying $\CP^1$ theory by $2\pi m\over p$.  Hence, as in the original $\CP^1$ theory, the Hilbert space includes distinct states with $-\pi < \theta \le \pi$, but the different superselection sectors are labeled by $-{\pi\over p} < \theta \le {\pi\over p}$.  The sum in the right hand side of \parfun\ can be interpreted as a sum over the $p$ different values of the background field in the same superselection sector. When the system includes additional charged fields $\varphi$ with $q=1$, only one of the $p$ different states labeled by $-\pi < \theta \le \pi$ in the same superselection sector is stable.  The others can decay through pair production and annihilation of the $\varphi$ particles as in \ColemanUZ.

\subsec{Various generalizations}

The construction above has a number of obvious generalizations.

We can repeat this construction with any two-dimensional nonlinear sigma model on any target space.  The total instanton number associated with any two-cycle can be constrained by adding an additional gauge field and a Lagrange multiplier as in \additer.

Another obvious generalization is to a higher dimensional spacetime.  Clearly, we can constrain the total winding number around any cycle.  Here $\lambda$ in \additer\ is a two form (as in BF-theories) and $\hat\theta=0$.

A somewhat more interesting generalization is to non-Abelian Yang-Mills theory in four dimensions.  The total instanton number can be constrained to be a multiple of $p$ by adding to the Lagrangian
\eqn\YMac{i\lambda (\nu - {p\over 2\pi} F^{(4)}) + i\hat \theta {1\over 2\pi} F^{(4)} }
where $\lambda $ is a Lagrange multiplier, the four form $\nu$ is the Pontryagin density normalized such that $\int \nu \in \Z$, and $F^{(4)}$ is a four form field strength of a three form gauge field normalized such that $\int F^{(4)} \in 2\pi \Z$.  The periodic variable $\hat \theta\sim \hat\theta + 2\pi$ takes the role of the ordinary $\theta$ angle which now has period $2\pi/p$.  With irrational $p$ or with the gauge group of $F^{(4)}$ being noncompact we have $\int F^{(4)}=0$ which leads to $\int \nu=0$.  Hence this theory does not have distinct superselection sectors labeled by $\theta$.\foot{ This does not lead to a solution to the strong $CP $ problem.  Instead, as in the discussion in section 2.1, the gauge theory of $F^{(4)}$ can have a background ``electric field'' which plays the role of $\theta$.  But since our system does not include dynamical charged 2-branes which couple to the three form gauge field, this background electric field cannot be screened and it is stable.}

The construction \additer\ might look contrived.  Therefore, we now present a more familiar theory which leads to the same effect, but the added topological degrees of freedom are different. We start with two scalars $z_{i=1,2}$ which are charged under a $U(1)$ gauge field $A$.  Normally, they are taken to have charge one, but we take them to have charge $p$.\foot{Such a system was considered from a more mathematical point of view in~\refs{\PantevRH\PantevZS-\CaldararuTC}.}  The scalars are subject to a potential with a minimum along the space $|z_1|^2 +|z_2|^2 = f_\pi^2$.  The low energy theory can easily be found.  If $z_2\not=0$ we can parameterize it by the gauge invariant (inhomogeneous) coordinate
 \eqn\hatzda{\Phi={z_1\over z_2}~, }
and we can solve
 \eqn\znpoe{z_2={f_\pi e^{i\alpha}\over \sqrt{1+ |\Phi|^2}}~ .}
The equation of motion of the gauge field sets
\eqn\Adefa{p A = \left(i {\bar \Phi \pa \Phi - \Phi \pa \bar
\Phi \over 2( 1+ |\Phi|^2)} - \pa \alpha\right) dx }
and leads to the Lagrangian \sigmalag.  The proper normalization of the gauge field \Adefa\ constrains the winding number as in \Iconst.

We see that this system is very similar to the example in section 2.2.  The local degrees of freedom are those of the $\CP^1$ model and the winding number is constrained by \Iconst.  However, globally these two systems are different. In section 2.2 we added a flat $U(1)$ gauge theory, while here it is a $\Z _p$ gauge theory.  This $\Z_p$ is the unbroken part of the underlying $U(1)$ gauge theory when $z_i$ get nonzero vevs.  As in section 2.2, we can add to this theory massive charged fields $\varphi$ which can induce transition between different, otherwise stable, ``$\theta$-vacua'' in the same superselection sector.  In section 2.2 the local degrees of freedom were the $\CP^1$ fields and the massive fields $\varphi$ with a global $U(1)$ symmetry (which was associated with the constrained $U(1)$ gauge theory).  Here we have the same variables but their Lagrangian has only a $\Z_p$ symmetry.

For use in later sections we consider now a supersymmetric version of this theory. There are two charged chiral superfields $z_{1,2}$ with a K\"ahler potential which includes an FI-term
\eqn\act{K= |z_1|^2e^{pV}+|z_2|^2e^{pV}-\xi V~.}
In four dimensions this theory is anomalous.  But this does not affect our classical analysis.  If $p,\ \xi >0$ this theory has a moduli space of supersymmetric vacua parameterized by
 \eqn\Mdef{\Phi={z_1\over z_2}~.}
The low energy effective theory is easily found by integrating out $V$.  Its equation of motion
\eqn\Veom{p(|z_1|^2+|z_2|^2) e^{pV}=\xi}
is solved by
\eqn\sol{V=-{1\over p} \log(1+|\Phi|^2) + {\rm chiral} + \overline {\rm chiral}~}
(compare with~\Adefa) and hence the low energy theory has
\eqn\acti{K_{eff}={\xi\over p}  \log(1+|\Phi|^2)~.}
The bosonic part of this Lagrangian is the same the theory discussed around \Adefa\ with
\eqn\fpixi{f_\pi^2 = {\xi \over p}~.}
As above, the quantization of the the topological charges in the low energy theory depends on the gauge group.  If the gauge group is $U(1)$, then the total wrapping number must be a multiple of $p$.  And if the gauge group is $\R$, the total wrapping number must vanish.

\newsec{Constraints on Supergravity Theories}

As a preparation for our discussion, we recall the well known fact that under the K\"ahler transformation (here and elsewhere in this section we set $ M_P^2 =1$)
\eqn\Kahlertr{K \to K +\Omega + \bar \Omega}
the superpotential $W$, the matter fermions $\chi^j$, the gauginos $\lambda$, and the gravitino $\psi_\mu$ transform as
\eqn\KahlertrW{\eqalign{ &W\to e^{-\Omega} W~, \cr
 &\chi^j \to e^{{1\over 4} (\Omega| - \bar \Omega|)} \chi^i~, \cr
 &\lambda \to e^{-{1\over 4} (\Omega| - \bar \Omega|)}\lambda~, \cr
 &\psi_\mu \to e^{-{1\over 4} (\Omega| - \bar \Omega|)}\psi_\mu ~,}}
where $\Omega|$ denotes the $\theta=\bar\theta=0$ component of $\Omega$.

\subsec{Theories with FI-terms}

We start by studying theories with FI-terms where
\eqn\Kwithxi{K=...-\xi V~.}
Here gauge transformations
\eqn\gaugetrn{V \to V + \Lambda + \bar\Lambda}
act as K\"ahler transformations with
\eqn\OmegaLambda{\Omega = - \xi \Lambda~.}
Therefore,~\KahlertrW\ means that the superpotential transforms as $W \to e^{\xi \Lambda}W$, which means that the gauge symmetry is an $R$-symmetry under which the superpotential has charge $-\xi$, and hence the supersymmetry coordinate $\theta$ has charge $-{\xi \over 2}$ and the gravitino $\psi_\mu$ has charge $ -{\xi \over 2}$.  If a superfield $\Phi^j$ has charge $q_j$ (i.e.\ it transforms as $\Phi^j \to e^{-q_j \Lambda}\Phi^j$), the fermion $\chi^j$ has charge $q_j +{\xi \over 2}$. (As a check, use Wess-Zumino gauge where the remaining gauge freedom is $\Lambda=i \alpha$ with real $\alpha$. Then use $\Omega = -i\xi \alpha$ in \KahlertrW\ to find the charges of the various fields.)

Is this compatible with the gauge symmetry of the problem?  Let us first assume that the gauge symmetry is $U(1)$ such that $\Lambda$ in \gaugetrn\ is identified with $\Lambda + 2\pi i$.  Then, charge quantization clearly implies that the scalars have integer charges; i.e.\ $q_j \in \Z $.  Examining the charges of the fermions we learn that the FI-term must be quantized\foot{Quantization of $\xi$ was considered by various people including \refs{\Wittenun,\Distlerun}.}
\eqn\xiqua{\xi= 2 N  \qquad {\rm with}
\qquad N \in  \Z~. }
If the gauge group is noncompact; i.e.\ it is $\R$ rather than $U(1)$, no condition like~\xiqua\ is required.

We would like to make three comments about these theories:
\item{1.} One might question the applicability of the condition \xiqua\ in supergravity when it is viewed as the low energy approximation of some more complete quantum gravity theory.  Then one might not want to consider coupling constants which are of order the Planck scale.
\item{2.} The theory includes charged fermions and one must make sure that all the anomalies are properly canceled.  For a recent discussion of anomalies in such theories see \refs{\ChamseddineGB\CastanoCI\BinetruyHH-\ElvangJK}.
\item{3.} To the best of our knowledge no example of theories satisfying~\xiqua\ were constructed in string theory.  This suggests that perhaps a deeper consistency condition might rule out some or even all of them.

\subsec{Theories with a nontrivial K\"ahler potential}

Next, we consider theories in which the K\"ahler form of the target space is not exact.  As a typical example, we study the $\CP^1$ model.  Following the discussion around \act\ we construct it in terms of a linear model of two charged chiral superfields $z_{1,2}$ with charge $p$.  Again, if the gauge group is $U(1)$, we have the condition \xiqua.  Combining this with \fpixi\ we have
\eqn\fpigra{f_\pi^2 = {\xi \over p} = {2N\over p}  \qquad {\rm with} \qquad p, N\in \Z .}

For $p=1$ equation \fpigra\ is the condition of Witten and Bagger \WittenHU.  However, we find that there is freedom in an arbitrary integer $p$.  As explained in section 2, it corresponds to a $\Z _p$ gauge theory.  In this case of a supergravity theory this $\Z _p$ symmetry is an R-symmetry.  Our fermions and gravitino transform under this $\Z _p$ symmetry and they play the role of the massive field $\varphi$ in section 2.  This allows the $p\not=1$ theory to be consistent, despite the fact that the condition of \WittenHU\ is not satisfied.

It is interesting to consider the $p\to \infty$ limit in which $f_\pi$ in \fpigra\ can be arbitrary.  In this limit the discrete symmetry $\Z _p$ becomes $\Z $.  One way to analyze this limit is to rescale $V$.  This effectively makes the gauge group noncompact. As we remarked after \xiqua, if the gauge group is $\R$, the FI-term $\xi$ is arbitrary and therefore $f_\pi$ is also arbitrary.   

One might object to using an effective Lagrangian with $f_\pi \sim M_{P}$.  However, as is common in string constructions, there are examples where the entire low energy theory is under control when various moduli change over Planck scale distances.

Finally, we would like to stress again that our condition \fpigra\ ignores various additional considerations.  For example, we might want to examine the consistency of the functional integral only for configurations which are close to solutions of the equations of motion.  In other words, when studying topological objects (like the instantons considered here) we might want to take into account the back reaction on the metric.  This can change the underlying spacetime and lead to different consistency conditions from the ones we discussed here.

\newsec{Recovering the Rigid Limit}

In this section we study theories in which the rigid limit $M_{P}\to \infty$ leads to supersymmetric field theories with a nonzero FI-term or with a nontrivial K\"ahler form.  (For that we restore the dimensions by appropriate factors of $M_P$.)  This allows us to connect with the results of~\KomargodskiRB\ and earlier references (see also the recent paper \KugoFS\ and references therein).

We start by considering an Abelian gauge theory with an FI-term $\xi$.  Equation \xiqua\ does not allow us to find a smooth rigid limit ($M_{P}\to \infty$) with finite $\xi$.  Therefore, the Abelian gauge group must be noncompact.

Furthermore, the fact that for finite $M_{P}$ the gauge group acts as an R-symmetry ($\theta$ has charge $-{\xi\over 2 M_{P}^2}$) puts interesting restrictions on the theory.  Consider first the rigid theory and assume that it has some matter fields $\Phi^j$ with gauge charges $q_j$.  When gravitational corrections are turned on the charges of the bosons can shift
\eqn\bosonsh{q_j \to q_j - r_j {\xi \over 2 M_{P}^2} +\CO\left({\xi^2 \over M_{P}^4}\right)}
with some order one constants $r_j$.  At the same time $\theta$ becomes charged and hence the superpotential must carry gauge charge $-{\xi\over M_{P}^2}$.  Consider a typical term in the superpotential
\eqn\typsup{W \supset\Phi^{j_1} \Phi^{j_2}\cdots}
Gauge invariance of the rigid limit demands
\eqn\rigd{q_{j_1} + q_{j_2} + ...=0}
and gauge invariance in the supergravity theory demands also
\eqn\supde{r_{j_1}+ r_{j_2} + ... = 2~.}
This means that the rigid theory has a global continuous R-symmetry under which $\Phi^j$ has R-charge $r_j$.  Equivalently, the supergravity theory has a global continuous non-R-symmetry under which $\Phi^j$ has charge $q_j$.\foot{More precisely, for this conclusion to be valid we need to make two assumptions. First, for the global symmetry to be nontrivial, we need to assume that at least one chiral superfield has $q_j\not=0$.  Second, we ignore the singular possibility of including terms in the Lagrangian in which the number of fields $\Phi^j$ diverges in the rigid limit like a power of $2M_{P}^2/ \xi$.}

We conclude that, if a theory with nonzero FI-term is to have a rigid limit, its gauge group must be $\R$ and the rigid theory should have a global R-symmetry \refs{\BarbieriAC\KalloshVE-\DvaliZH}.  Furthermore, the supergravity theory has a continuous global symmetry \KomargodskiPC.

We should emphasize that this conclusion about the presence of a global continuous symmetry follows from our assumption in this section that $\xi\over M_{P}^2$ is parametrically small.  In the context of the discussion in section 3 we can easily find supergravity theories with no global symmetries.  For example, let $\xi = 2M_{P}^2 $ (i.e.\ $N=1$ in \xiqua) and consider a theory with two chiral superfields $\Phi_\pm$ with gauge charges $\pm 1$.  Then, the superpotential $W=\Phi_-^2(a_0 + a_1 (\Phi_+\Phi_-) +  a_2 (\Phi_+\Phi_-)^2 + ...)$ with come constants $a_i$ carries the desired gauge charge without additional global symmetry.

Next we discuss theories with a nontrivial K\"ahler form.  Here we want to consider the rigid limit $M_{P} \to \infty$ with fixed $f_\pi$.  Constructing such theories using gauged linear models we can use the result above that the gauge group must be $\R$ and the rigid theory should have a global continuous R-symmetry.  Alternatively, we can use \fpigra\ and take the $M_{P}\to \infty$ limit together with $p\to \infty$.

We conclude that if we are willing to consider supergravity theories with continuous global symmetries, not only can we have theories with FI-terms, we can also have sigma models with nontrivial K\"ahler forms \KomargodskiRB.  An alternate way to construct these supergravity theories is to consider the ``new minimal'' auxiliary fields of supergravity~\refs{\AkulovCK,\SohniusTP}.\foot{The authors of \LambertDX\ argued that globally this theory differs from the more standard ``old minimal'' theory.}  This amounts to gauging the R-multiplet rather than the Ferrara-Zumino multiplet~\KomargodskiRB\ (see also~\refs{\DienesTD,\KuzenkoYM}).

We should stress, however, that a consistent theory of quantum gravity cannot have any global continuous symmetries.  Therefore such supergravity theories cannot be realized~\refs{\KomargodskiPC,\KomargodskiRB}.  This is an example of a point we have made a number of times above that our classical considerations lead only to necessary conditions and it is quite possible that additional, more subtle considerations put further restrictions on the theories studied here.

\bigskip
\centerline{\bf Acknowledgements}
We would like to thank Z.~Komargodski and M.~Rocek for participation in early stages of this project and for many useful comments.  We have also benefitted from useful discussions with O.~Aharony, N.~Arkani-Hamed, J.~Distler, M.~Douglas, D.~Freed, J.~Maldacena, G.~Moore, S.~Shenker, Y.~Tachikawa, and E.~Witten. The work of  NS was supported in part by DOE grant DE-FG02-90ER40542. Any opinions, findings, and conclusions or recommendations expressed in this material are those of the author(s) and do not necessarily reflect the views of the funding agencies.

\listrefs
\end